\documentclass[conference]{IEEEtran}
\IEEEoverridecommandlockouts

\usepackage{orcidlink}
\usepackage{cite}
\usepackage{caption}
\usepackage{amsmath,amssymb,amsfonts}
\usepackage{algorithmic}
\usepackage{graphicx}
\usepackage{textcomp}
\usepackage{xcolor}
\usepackage{comment}
\usepackage{array}
\usepackage{tabularx}
\usepackage{booktabs} % For formal tables
\usepackage{listings}
\usepackage{caption}
\usepackage{mathtools} % For multiline equations
\usepackage[T1]{fontenc}
\usepackage{units}
\def\BibTeX{{\rm B\kern-.05em{\sc i\kern-.025em b}\kern-.08em
    T\kern-.1667em\lower.7ex\hbox{E}\kern-.125emX}}
\begin{document}

\title{DIDChain: Advancing Supply Chain Data Management with Decentralized Identifiers and Blockchain}

\author{\IEEEauthorblockN{
Patrick Herbke \orcidlink{0000-0001-9649-2975}\IEEEauthorrefmark{1}, 
Sid Lamichhane \orcidlink{0009-0000-5845-7060}\IEEEauthorrefmark{1}, 
Kaustabh Barman \orcidlink{0009-0001-9782-8917}\IEEEauthorrefmark{1}, 
Sanjeet Raj Pandey \orcidlink{0009-0007-9525-7587}\IEEEauthorrefmark{1}, 
Axel K\"upper \orcidlink{0000-0002-4356-5613}\IEEEauthorrefmark{1} \\ 
Andreas Abraham \orcidlink{0000-0002-4163-9113}\IEEEauthorrefmark{2}, 
Markus Sabadello \orcidlink{0000-0003-3603-0239}\IEEEauthorrefmark{3}
}
\IEEEauthorblockA{\IEEEauthorrefmark{1}Service-centric Networking,
Technische Universit\"at Berlin, Berlin, Germany\\
\{p.herbke, lamichhane, kaustabh.barman, s.pandey, axel.kuepper\}@tu-berlin.de}
\IEEEauthorblockA{\IEEEauthorrefmark{2}ValidatedID S.L, Barcelona, Spain, 
andreas.abraham@validatedid.com}
\IEEEauthorblockA{\IEEEauthorrefmark{3}Danube Tech GmbH, Vienna, Austria, 
markus@danubetech.com}}

\maketitle

\begin{abstract}
Supply chain data management faces challenges in traceability, transparency, and trust. These issues stem from data silos and communication barriers. This research introduces DIDChain, a framework leveraging blockchain technology, Decentralized Identifiers, and the InterPlanetary File System. DIDChain improves supply chain data management. To address privacy concerns, DIDChain employs a hybrid blockchain architecture that combines public blockchain transparency with the control of private systems. Our hybrid approach preserves the authenticity and reliability of supply chain events. It also respects the data privacy requirements of the participants in the supply chain. Central to DIDChain is the cheqd infrastructure. The cheqd infrastructure enables digital tracing of asset events, such as an asset moving from the milk-producing dairy farm to the cheese manufacturer. In this research, assets are raw materials and products. The cheqd infrastructure ensures the traceability and reliability of assets in the management of supply chain data. Our contribution to blockchain-enabled supply chain systems demonstrates the robustness of DIDChain. Integrating blockchain technology through DIDChain offers a solution to data silos and communication barriers. With DIDChain, we propose a framework to transform the supply chain infrastructure across industries.
\end{abstract}


\section{Introduction}
The global supply chain, a complex network of interconnected processes and entities, is crucial to the economic and social fabric of societies~\cite{koberg2019systematic}. Supply chains move goods from producers to consumers, involving manufacturers, suppliers, and retailers~\cite{kim2017impact}. Supply chain efficiency directly influences the cost, quality, and timeliness of products~\cite{AZADI2015274}. However, the complex nature of supply chains presents challenges in managing and optimizing these processes. These challenges include traceability, transparency, and trust~\cite{khan2021state}. 

Traceability, the ability to track and trace goods from origin to consumer, is crucial to ensuring the authenticity and safety of the product~\cite{SUNNY2020106895}. Transparency, the visibility of supply chain operations to all stakeholders, builds trust and facilitates informed decision-making~\cite{sodhi2019research}. Trust in the integrity and reliability of all actors is fundamental for global supply chains~\cite{kwon2004factors}.

\textbf{Challenges.} Supply chain data management (SCDM) is central to these challenges~\cite{hazen2014data}. The vast amount of information in supply chains necessitates sophisticated data management systems~\cite{wang2016big}. SCDM systems must support decision-making, improve efficiency, and enhance supply chain resilience~\cite{sayogo2015challenges}. Effective SCDM faces challenges such as heterogeneity of data sources, the rapid pace of technological change, the need for real-time data processing, and the complexity of integrating data across different systems and platforms~\cite{ageron2020digital}. Supply chain variability and dynamic market conditions also challenge data accuracy and reliability~\cite{barbosa2018opportunities}.

\textbf{Contribution.}
This paper introduces DIDChain, a novel approach to SCDM leveraging blockchain technology~\cite{saberi2019blockchain}, Decentralized Identifiers (DIDs)~\cite{reed2020decentralized}, and the InterPlanetary File System (IPFS)~\cite{naz2019secure} for data traceability and trust. Comparing DIDChain with a similar smart contract-based approach~\cite{DBLP:conf/ithings/WesterkampVK18} highlights our framework's ability to maintain high levels of traceability and reliability in supply chain data management without gas costs and the complexities associated with verifiable credentials. DIDChain incurs transaction fees specific to the cheqd blockchain, detailed in our cost analysis.

\section{Related Work}
In SCDM, the integration of blockchain, DIDs, and IPFS is revolutionizing how entities verify and trace assets across global supply chain networks~\cite{azzi2019power}. DIDs facilitate digital identities, enabling secure and verifiable interactions between entities without centralized oversight~\cite{malik2019trustchain}. Blockchain technology provides an immutable ledger for recording supply chain events and asset movements, ensuring traceability and trust~\cite{kamilaris2019rise}. IPFS addresses the scalability challenges inherent in blockchain technology by offering a decentralized solution for efficient data storage and retrieval~\cite{8946164}.

Salah et al.~\cite{8718621} propose using the Ethereum blockchain and smart contracts to improve agricultural supply chain traceability, eliminating centralized authority. Their system ensures integrity and reliability by recording supply chain events on a blockchain and linking them to IPFS. Our research avoids the gas costs associated with Ethereum smart contracts. Unlike Salah's system, DIDChain leverages the cheqd blockchain, which uses a Tokenomics model with fixed transaction fees~\cite{cheqd2024price}.

Wang et al.~\cite{wang} introduce a smart contract-based framework that uses Hyperledger technology to improve traceability within the agricultural food supply chain. By incorporating blockchain functionalities, Wang's framework addresses issues in traditional supply chains, including inefficiencies, insufficient transparency, and compromised data integrity. The research by Wang et al. enables data sharing and removes informational barriers between supply chain participants within Hyperledger technology. In contrast, our study presents a framework that can be applied to various blockchains.

Westerkamp et al.~\cite{westerkamp2020tracing} introduce a blockchain-based system using token recipes to trace goods within supply chains on the Ethereum Virtual Machine. Their method creates nonfungible tokens for each batch of products, linking them to their components, and preserving origin information. DIDChain, on the contrary, uses DIDs and IPFS for data management. Westerkamp’s approach focuses on manufacturing processes, but incurs higher costs and lacks integrated digital identity management.

Several challenges remain in current research, especially in relation to the transparency and trust of SCDM. This paper introduces an SCDM framework that improves traceability, transparency, and trust using blockchain, DIDs, and IPFS. Our proposed DIDChain framework facilitates secure, efficient, and reliable supply chain operations, addressing existing limitations. However, our evaluation shows that DIDChain faces challenges in real-world scalability and economic feasibility, which require further improvements.

\section{Preliminaries} 
We will explore how DIDChain leverages blockchain, DIDs, and IPFS in supply chain events. 

\subsection{Tracking vs. Tracing in Supply Chain Data Management}
In SCDM, the terms tracking and tracing assets refer to raw materials or products and have distinct meanings~\cite{wu2012peertrack}. 

\textit{Tracking} monitors current and future states of supply chain events, such as the shipment of raw materials from one manufacturer to another. This process allows stakeholders to anticipate and plan for future transactions by viewing all supply chain events. 

\textit{Tracing} involves a global view of the entire history of a product or material, including obtaining references to the used compartments if it is a product. This comprehensive view of an asset's history and current status is crucial for ensuring product authenticity and safety, as well as for building trust and facilitating informed decision-making.

DIDChain improves traceability by providing a framework for tracing the history of assets from their origin to current status. The trace includes a complete record of an asset, detailing ownership changes and status updates, such as production, manufacturing, and shipment events. Implementing a prototype based on DIDChain ensures that every change in ownership is recorded on the blockchain, ensuring traceability across the supply chain. DIDChain's use of a public and permissionless blockchain provides everyone with access to the data essential for tracing. Public and permissionless blockchain makes it easy for supply chain entities to track goods after they leave ownership.

\subsection{Supply Chain Events}
Supply chain events are stages within a supply chain. DIDChain models these stages through four atomic events:
\begin{itemize}
    \item \textit{Producing.} Initial creation or transformation of raw materials (e.g. extracting milk from dairy cows).
    \item \textit{Shipping.} Transporting assets, ensuring the integrity and safety of the products throughout their journey.
    \item \textit{Receiving.} Assets are received after shipment and documented for quality control and legal purposes.
    \item \textit{Manufacturing.} Transforming raw materials into finished products (e.g. making cheese from milk and yeast).
\end{itemize}

\begin{figure}[!h]
    \centering
    \includegraphics[width=\linewidth]{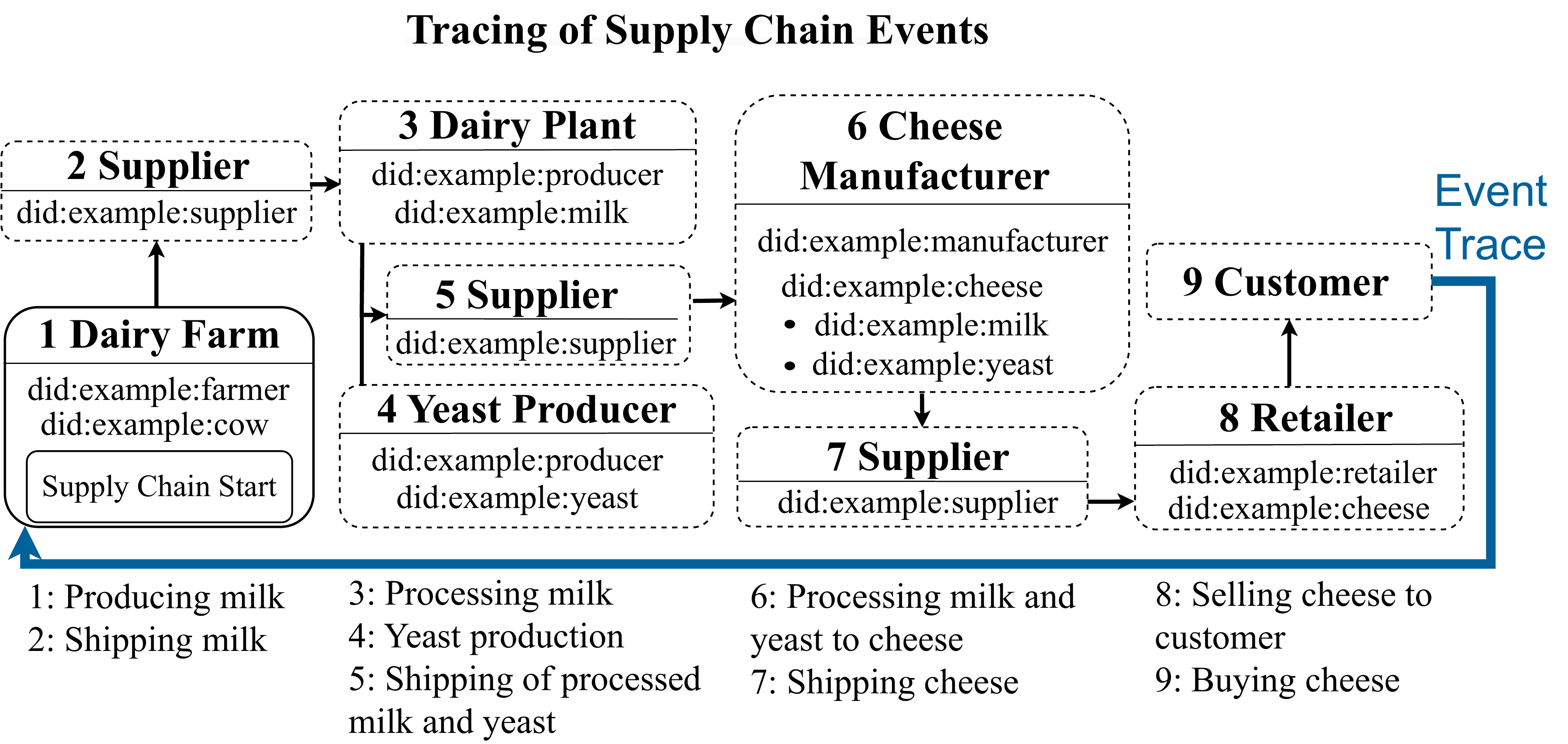}
    \caption{Supply chain with stages and stakeholder interactions. The stages are indexed from (1) production to (9) consumer engagement, including unique DIDs assigned to each entity and asset.}
    \label{fig:supply-chain-process}
\end{figure}

Figure~\ref{fig:supply-chain-process} illustrates these atomic supply chain events, with each entity from \texttt{dairy farm to customer} uniquely identified by a DID. The DIDChain framework records events from production to purchase, embedding the DIDs of raw materials and products in DID Documents. In cheese production, producers record the unique identifier (DID) of the milk from the dairy farm in the milk DID Document, enabling them to trace the cheese back to its origin.

\subsection{Decentralized Identifiers (DIDs) and DID Documents}
In contrast to traditional federated identifiers, DIDs function autonomously from central registries, identity providers, and certification authorities. This autonomy allows the owner of a DID to authenticate the ownership directly, bypassing the need for external validation~\cite{muhle2018survey}. DIDs are Uniform Resource Identifiers that link a subject and public key to a DID Document, facilitating trustworthy interactions. Each DID Document contains public keys, verification methods, and services, empowering the DID Controller to demonstrate control and resolve resources associated with the subject. The structure of a DID looks like \texttt{did:example:123456789abcdefghi}, characterized by \texttt{did} as the scheme, \texttt{example} indicating the DID method, and \texttt{123456789abcdefghi} representing the unique identifier, generated by the method-specific DID registry~\cite{brunner2020did}. A DID Method is a specification for how DIDs are created, updated, and resolved in a decentralized manner. Our methodology uses DIDs for asset tracking, associating each supply chain asset with a unique DID, potentially through QR codes, to ensure complete traceability from consumer to origin.

The proposed DIDChain framework conceptualizes DID Documents as digital twins of physical assets, encompassing products and raw materials. These digital twins encapsulate asset information, including ownership, characteristics, functions, and lifecycle data, mirroring the physical asset~\cite{singh2021digital}. While modifications in digital twins can influence physical assets, this research's primary objective is to mirror the asset's state, not to enable bidirectional changes between digital and physical entities. Creating a digital twin is related to generating a DID, which serves as the unique identifier for the digital asset. The incorporation of digital twins into a blockchain necessitates the compilation of detailed asset information, thereby enhancing secure digital interactions and the verification of asset statuses.

\subsection{Blockchain}
Blockchain technology, grounded in Distributed Ledger Technology (DLT), enhances SCDM by offering a decentralized, transparent, and immutable framework. In the scope of our research, blockchain technology ensures that supply chain events are recorded in a manner that is not centralized, allowing for a wide distribution of data across multiple entities without the need for a central authority. A blockchain achieves transparency by being open and accessible, making all transactions and events visible to all network participants. The immutability of blockchains ensures that recorded events cannot be altered, providing a permanent record~\cite{lim2021literature}. However, blockchain technology faces challenges in scalability, especially in data storage and processing~\cite{kouhizadeh2021blockchain}. DIDChain combines IPFS with blockchain to address these challenges. Our framework integrates error correction mechanisms to preserve the integrity of the blockchain. DIDChain manages incorrect entries through DID document versioning, allowing the appending of corrected versions while maintaining a clear audit trail.

\subsection{InterPlanetary File System (IPFS)}
IPFS addresses blockchain’s scalability challenges in SCDM by decentralizing data storage and eliminating reliance on centralized servers. IPFS ensures data immutability by assigning a unique content-based hash to each file, referring to metadata for each supply chain event. This off-chain metadata includes asset descriptions, owner identities (DIDs), and other relevant data. The SHA-256 cryptographic hash transforms metadata $x$ into a unique 256-bit hash $H(x)$, ensuring any alteration results in a new hash. This hash is recorded on-chain within DID Documents, combining blockchain for immutable event logging and IPFS for efficient data storage, thus ensuring integrity and transparency in supply chain events~\cite{Naz2019ASD}.

\section{Concept and Implementation}
The DIDChain framework uses blockchain, DIDs, and IPFS to improve SCDM. By documenting each supply chain event on the blockchain, we ensure transparency and trust in tracing assets from their origin to the consumer. DIDs provide each asset or owner with a unique and verifiable identity, ensuring authenticity and traceability. IPFS addresses data scalability, efficiently managing asset metadata such as production timestamps, batch numbers, and source materials, enriching the digital metadata of assets for comprehensive tracing and verification.

The DIDChain framework ensures a seamless, verifiable information flow, with DIDs stored on the blockchain and asset metadata stored off-chain.

\subsection{Roles}
In the DIDChain framework, Producers, Suppliers, Manufacturers, Retailers, and Customers seamlessly manage digital assets. These roles facilitate the entire supply chain process, from raw material production to the distribution of finished products to end consumers.

\begin{itemize}
    \item \textit{Producers} are at the origin of the supply chain and responsible for the initial production of raw materials. Producers create DIDs representing materials and initiating the supply chain process.
    \item \textit{Suppliers} provide manufacturers with raw materials, ensuring the availability of essential inputs for the production process. Suppliers hold and update DIDs for assets awaiting distribution.
    \item \textit{Manufacturers} transform raw materials into finished products. They update each DID corresponding to an asset as it progresses through the supply chain. When these assets contribute to the creation of a new product, manufacturers generate new DIDs to reflect this transition.
    \item \textit{Retailers} are intermediaries in the supply chain, purchasing products from manufacturers and distributing them to end consumers. They continue to hold and update DIDs as they handle the products.
    \item \textit{Customers} are the end-users of manufactured products. They verify products or materials by resolving DIDs to ensure origin and authenticity.
\end{itemize}

\subsection{Key Management}
The DIDChain framework maintains cryptographic key management and the generation of DIDs through the \texttt{did:cheqd} method as a component to establish secure and verifiable digital assets. The cheqd credential service facilitates the creation of DIDs and updates on asset-related DID Documents along with supply chain events~\cite{cheqd2024credservice}. Cheqd APIs offer the creation, management and secure storage of DIDs, ensuring the integrity and authenticity of digital assets within the supply chain. Cryptographic key management, including the handling of public and private keys, is integral to maintaining the security and verifiability of digital identities. The~\texttt{did:cheqd} method operates on the Cosmos blockchain~\cite{CosmosNetwork}, hereafter referred to as the cheqd blockchain. 

Cheqd offers an API for the credential service~\cite{cheqd2024credserviceapi} to manage public and private keys related to DIDs within the cheqd infrastructure. The credential service API enables the creation and management of DIDs and DID Documents. The cheqd API provides a range of endpoints that allow the creation of identity (asset) key pairs, the import of existing key pairs, and the retrieval of key pairs. The keys are stored on the cheqd infrastructure, ensuring their integrity and availability for subsequent operations. The cheqd API also includes endpoints for managing DID Documents, such as creating, updating, importing, and deactivating DIDs. These operations are essential for maintaining the lifecycle of assets and related DIDs within the cheqd ecosystem. The \texttt{/did/list} endpoint enables users to fetch DIDs associated with an account. The \texttt{/did/search/\{did\}} endpoint allows for the resolution of a DID Document.

Various solutions enable generalized DID operations beyond a single DID method, such as \texttt{did:cheqd}, currently used by DIDChain. The DID Resolution specification~\cite{w3cccg2024didresolution} of the W3C Credentials Community Group defines a generic interface to resolve DIDs. The DID Registration specification of the Decentralized Identity Foundation (DIF)~\cite{dif2024didregistration} specifies the creation, updating, and deactivation of DIDs. DID operations are universally applicable for various DID methods, including \texttt{did:cheqd}, \texttt{did:ebsi}, \texttt{did:indy}, and \texttt{did:ion}. Thus, a future version of DIDChain could be DID method-agnostic. The adoption of \texttt{did:cheqd} for DIDChain is due to its robust DID management features and the cheqd API support for DID operations, ensuring secure and authentic digital assets. Integrating the \texttt{did:cheqd} method with IPFS and tracing detailed DID Document histories enhances transparency and immutability in the supply chain.

The DID Registration specification outlines architectures for managing a DID’s private keys. DID registration options include the \texttt{internal secret mode}, where a hosted service such as the cheqd Credential Service or Godiddy.com~\cite{godiddy2024} manages the private keys. Another mode is \texttt{client-managed secret mode}, where private keys remain client-side, separate from DID operations. In \texttt{client-managed secret mode}, control over a DID remains independent of intermediaries. Clients can use key management systems, such as hardware tokens and secure elements.

\subsection{DID Document Versioning and History Linking}
\begin{figure*}[!htbp]
    \centering
\includegraphics[width=0.7\linewidth]{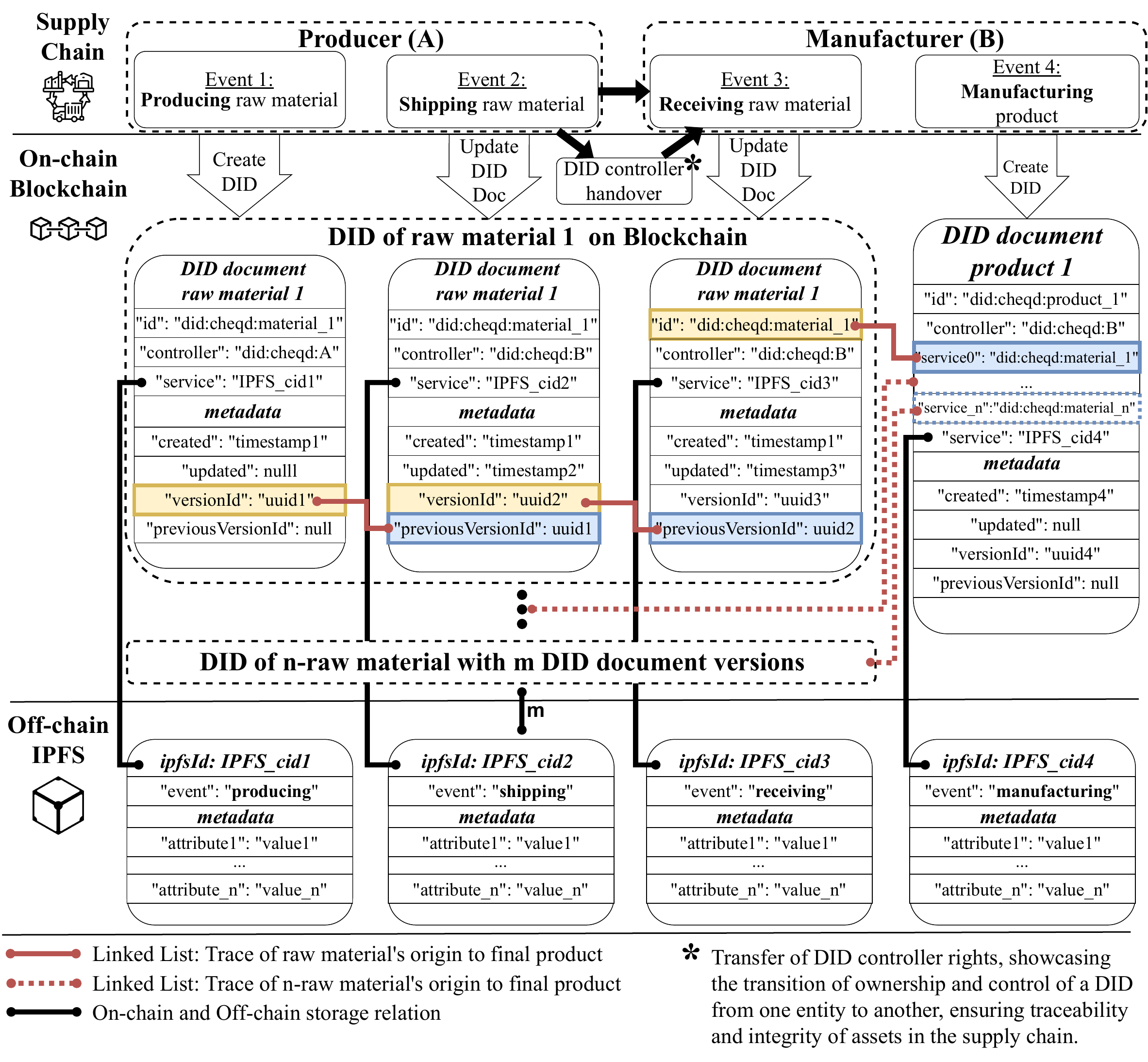}
    \caption{Integration of Blockchain and IPFS in Supply Chain: Showcasing the lifecycle of assets through DID Documents as digital twins, from raw material to final product, enhancing traceability and verification.}
    \label{fig:process_detail}
\end{figure*}

DID Document versioning and historical linkage are essential in our blockchain-based DIDChain framework, enhancing asset integrity and traceability throughout the supply chain. Each raw material and product has a unique DID and an associated DID Document on the cheqd blockchain, managed by the DID controller. DIDs and DID Documents are  immutable digital identifiers that enable tracing and verifying assets through supply chain events. The controller updates DID Documents with each supply chain event, maintaining an accurate and verifiable history of ownership and status changes. Active participation of supply chain participants is required to ensure accurate and comprehensive data recording. Figure~\ref{fig:process_detail} illustrates the details of our supply chain framework.

% TODO: Add sent paper!
A DID Document’s history provides an immutable audit trail for each asset in the supply chain. DID Documents are irreversibly updated with each event, reflecting asset transfers and attribute changes. The cheqd blockchain maintains an on-chain, immutable, and verifiable record of all DID Document versions, including metadata such as timestamps and version links, ensuring trustworthiness. The historical access to all assets confirms the provenance and authenticity of all assets, deterring counterfeits. Our framework includes the DIDs of the raw materials in each product's services section to enable traceability. The immutable history of DID Documents underpins the handover of DID Controller rights, maintaining a coherent digital representation of asset ownership. Figure~\ref{fig:process_detail} illustrates the DID document versioning through metadata fields such as \texttt{created}, \texttt{updated}, and \texttt{VersionId}. The red arrows trace the transition of DID Documents from raw materials to products within the supply chain.

\subsection{DID Ownership} 
As assets traverse the supply chain, ownership of associated DID and DID Document rights is shifted between entities, reflecting changes in control and management. This process, initiated by supply chain events, involves updating the DID Document with the new owner's public keys, thereby synchronizing the asset's digital and physical identities. The immutable and permanent record of these ownership changes on the blockchain establishes a transparent and unalterable audit trail, improving the integrity and trustworthiness of the supply chain and the DIDChain framework.

In the DIDChain framework, the DID Controller has the authority to modify a DID Document, ensuring the integrity and traceability of assets within the supply chain. The selected DID method, \texttt{did:cheqd}, details the authorization process for a DID Controller~\cite{cheqd_did_method}. The controller property in a DID Document lists one or more DIDs, and any verification methods for those DIDs are considered authoritative. In DIDChain, proofs satisfying the verification methods in the DID Document are equivalent to proofs from the DID subject. The DID Controller can change the DID Document and prove control over the DID through these authoritative methods.

The transfer of DID Controller rights synchronizes physical assets with their digital counterparts on the blockchain, triggered by supply chain activities such as shipping or manufacturing. This transfer involves passing the DID and its DID Document rights to a new entity, updating the DID Document with the new owner’s public keys, and aligning digital and physical ownership. These updates are permanently recorded on the blockchain, preserving asset integrity and traceability.

If malicious actors gain access to a private key to a DID controller, they could corrupt the supply chain by unauthorized modifications. Figure~\ref{fig:process_detail} illustrates the~\texttt{DID Controller handover} between the producer and the manufacturer, with arrows indicating the transfer of ownership and the updated 'controller' fields within the DID Documents.

\subsection{Revocation}
Revocation within the DIDChain framework safeguards the integrity and reliability of SCDM. It is crucial when a DID is compromised, linked to a discontinued entity, or when the digital asset is no longer relevant. Deactivating DIDs and updating a DID Document’s status to \texttt{withdrawn} are essential actions in such cases. Deactivating a DID blocks subsequent transactions or operations under that identifier, preserving system security and trustworthiness. Updating DID Documents to indicate a \texttt{withdrawn} status informs all stakeholders that the material or product is no longer part of active supply chain operations.

Deactivating DIDs may not be suitable in certain supply chain situations, such as when an asset is permanently lost, suffers irreparable damage, or fails to meet required standards. In a circular economy, assets considered unusable can become compartments within other products, integrating them again into the supply chain. Therefore, revocation necessitates a thoughtful approach to DID deactivation, considering their reuse potential and the implications for the asset's historical tracing and traceability. Additionally, revocation is crucial when keys are lost or compromised, enabling the supply chain to maintain operations without needing re-authentication.

\subsection{Data Integrity and IPFS Linkage}
The DIDChain framework addresses data integrity in IPFS by leveraging the authority of the current DID Controller. The framework asserts that signatures on IPFS-stored data are redundant when (1) the DID Document's service endpoint directly references the IPFS data, (2) the IPFS link intrinsically secures the data's integrity, and (3) such linkage is exclusively authorized by the current DID Controller. The system is vulnerable to unauthorized changes if the controller's private key is compromised. DIDChain can be extended with two-factor authentication for enhanced security. Signatures are only required when the authoritative issuer of the event data differs from the DID that embeds the IPFS link in its document. This delineation allows for using JSON-LD for schema linking without categorizing the data as a Verifiable Credential, optimizing the framework's design for enhanced data integrity and management efficiency without the complexities of verifiable credentials administration.

\subsection{Implementation}
\begin{figure}[!ht]
    \centering
    \includegraphics[width=.9\linewidth]{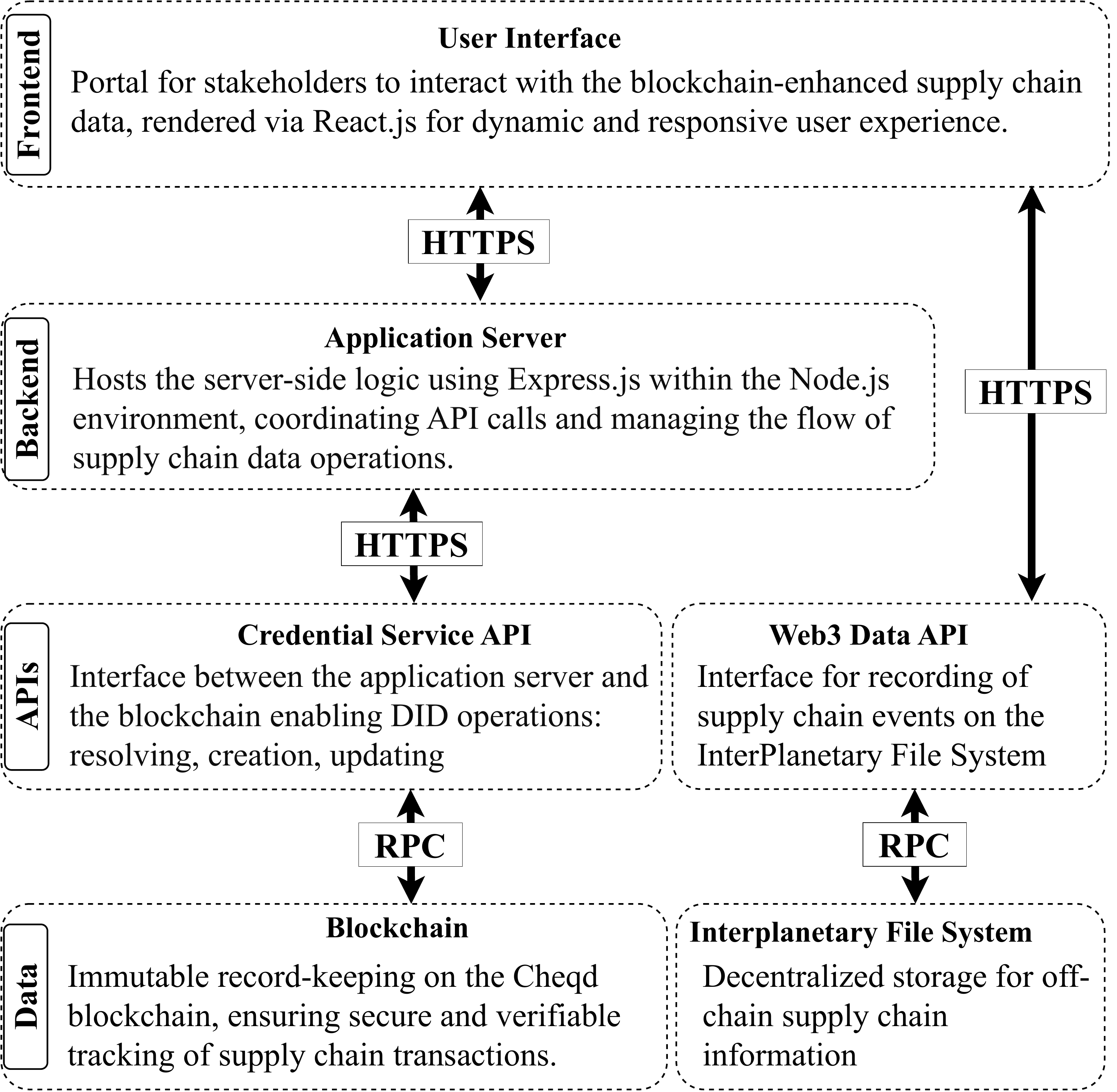}
    \caption{Detailed architecture of the SCDM framework showcasing the integration of frontend technologies with blockchain and IPFS via secure APIs.}
    \label{fig:architecture}
\end{figure}

The DIDChain framework, illustrated in Figure~\ref{fig:architecture}, employs an architecture that integrates the cheqd blockchain, IPFS, and security features offered by HTTPS and RPC protocols. In implementing DIDChain's hybrid blockchain architecture, sensitive data is managed within permissioned ledger components that enforce privacy controls, while the immutable public ledger facilitates transparency and broad verification processes.

To integrate DIDChain into existing supply chain systems, a compatibility assessment of the system with blockchain technology is essential, accompanied by a detailed plan to migrate existing data to the DIDChain framework. Using cheqd and Moralis APIs facilitates seamless communication while implementing two-factor authentication enhances security. This integration ensures secure, transparent, and efficient supply chain operations.

Step-by-step implementation:
\begin{enumerate}
    \item \textbf{Setup}: Configure the cheqd blockchain and IPFS.
    \item \textbf{Data Management}: Store sensitive data within permissioned ledger components and use IPFS for efficient data storage.
    \item \textbf{Security Integration}: Implement HTTPS and RPC protocols for secure data transmission.
    \item \textbf{Event Documentation}: Record supply chain events in the public ledger for transparency.
    \item \textbf{DID Management}: Create and manage DIDs for assets, ensuring traceability and integrity.
    \item \textbf{Privacy Controls}: Enforce privacy through permissioned ledgers, balancing transparency and confidentiality.
\end{enumerate}

The frontend enables the documentation and tracing of asset supply chain events by interacting with IPFS through Moralis's Web3 Data API~\cite{moralis}. The backend facilitates communication between the frontend and cheqd via the Credential Service API, ensuring API key security.

The cheqd blockchain and IPFS collectively serve as the data layer, with IPFS enhancing scalability and minimizing on-chain storage. Cheqd’s design for DID Document versioning includes automatic timestamping and linking to previous versions. These built-in features ensure public, permanent, and immutable versioning information of assets stored in DID Document metadata, enabling transparent and trustworthy asset tracing.

The choice of cheqd and Moralis API was driven by cheqd’s DID management capabilities and Moralis’s IPFS integration, aligning with DIDChains' objectives. Cheqd can be replaced in future research to explore alternative blockchains and APIs that offer distinct advantages or address specific development challenges.

The GitHub repository~\cite{sid030sid2024} contains an implementation of the DIDChain framework and its evaluation. The repository is carefully organized and includes detailed documentation.

\section{Evaluation}
We assess the system's integration with the cheqd blockchain technology, focusing on transactional efficiency, cost-effectiveness, and constraints on the DID Document capacity as core performance indicators for real-world adoption. Empirical analysis is supported by a data set that simulates actual supply chain events, providing a basis for benchmarking the DIDChain framework against real-world operational requirements. Additionally, we explore the implications of the system's performance on trust establishment within SCDM processes. Our findings are compared with existing benchmarks to establish the operational efficacy and economic viability of the DIDChain framework.

The performance evaluation uses JavaScript in the Node.js environment \texttt{(version 20.8.1)}, with Axios \texttt{(version 1.6.7)} as the HTTP client, ensuring efficient request handling. This configuration guarantees precise and reliable testing results. This setup provides a robust and efficient platform for the execution of the tests, ensuring that the limitations of the testing environment do not compromise the results obtained. The test results are recorded and stored in CSV files locally, allowing independent analysis. The hardware configuration includes a system running Windows 11, equipped with an AMD Ryzen 7 7735HS processor operating at 3.20 GHz and 16 GB of RAM. The cheqd blockchain testnet is used during testing to simulate real-world blockchain interactions.

\subsection{Economic Feasibility}
We analyzed the costs to assess the economic feasibility of DIDChain. Excluding the costs of maintaining off-chain data in IPFS, stakeholders' primary expenses are incurred through the cheqd blockchain. The cheqd network has fixed blockchain operation costs~\cite{cheqd_proposal_30}. As of March 5th, one cheqd token (CT) costs $\$0.117$~\cite{BigDipperCheqd2024}. DID Creation is a blockchain write operation that costs 50 CT while updating a DID Document requires an operation cost of 25 CT~\cite{cheqd_fee_structure}. Table~\ref{tab:cost_per_stakeholder} summarizes the cost of stakeholder interaction with the cheqd blockchain.

\begin{table}[ht]
\renewcommand{\arraystretch}{1.2}
\centering
\caption{Transaction costs for stakeholder activities within the DIDChain framework as of 05 March 2024. This table itemizes the costs incurred by each stakeholder for the creation and update operations of DIDs, measured in Cheqd Tokens (CT) and US Dollars (\$), predicated on the valuation of one CT at \$0.117. The DID Operation of the manufacturer depends on \texttt{n} numbers of compartments per product.}

\begin{tabular}{p{0.12\columnwidth} p{0.22\columnwidth} p{0.23\columnwidth} p{0.24\columnwidth}}
\toprule
\textbf{Stakeholder} & \textbf{Actions} & \textbf{DID Operation} & \textbf{Cost (CT | \$)} \\
\midrule
Producer & Produces, ships material & 1$\times DID_{\text{Creation}}+1 \times DID_{\text{Update}}$ & 75 | 8.78\\
Supplier & Receives, ships material/product & $2\times DID_{\text{Update}}$ & 50 | 5.85 \\
Manufact. & Receives materials, manufactures, ships product & $1\times DID_{\text{Creation}}+(1+n)\times DID_{\text{Update}}$ & \text{50}$+(1+n)\times$ 25 | $5.85+(1+n)\times 2.93$\\
Retailer & Receives, ships product & $2\times DID_{\text{Update}}$ & 50 | 5.85\\
Customer & Receives product & $1\times DID_{\text{Update}}$ & 25 | 2.93\\
\bottomrule
\end{tabular}
\label{tab:cost_per_stakeholder}
\end{table}

Considering the documentation of manufacturing a car with 30,000 compartments, the total cost in CTs for the manufacturer is calculated as follows: 

\begin{equation}
    \begin{aligned}
    \text{Total Cost}_{\text{CT}} = & 1\times \text{Cost}_{\text{DID Creation}}+(1+ N_{\text{Compartments}}) \\ & =\times \text{Cost}_{\text{DID Update}} \\
    &= 50 + (1 + 30,000) \times 25 \\
    &= 750,075 \, \text{CT},
    \end{aligned}
\end{equation}
\begin{equation}
\text{Total Cost}_{\text{USD}} = 750,075\text{ CT} \times \unitfrac[0.117]{\$}{CT} = \$87.785,78
\end{equation}

The total cost of documenting a car with 30,000 compartments, including DID Creation and DID Document updates, is \$87.785,78. The cost of documenting a product's manufacturing can be lowered if manufacturers do not document the receiving of each used compartment and store the information about receivables in the IPFS file of the producing event. Omitting documentation of asset reception events in supply chains could fix the cost at 75 currency units per product, regardless of compartment usage, but it decreases traceability.

Compared to the research by Westerkamp et al.~\cite{westerkamp2020tracing}, creating a DID for an asset is equivalent to creating a batch with one item to ensure the same level of traceability as provided by DIDChain. Creating one batch costs 92,634 gas plus a factor of 39,340 gas multiplied by the number of compartments. This gas calculation translates to \$25.77 plus \$10.94 multiplied by the number of compartments, resulting in a total cost of \$328,225.77 for a car with 30,000 compartments. The cost analysis of the DIDChain framework, which does not include IPFS off-chain data maintenance, is \$87,785.78. The pricing structure for the IPFS is as follows: the service is complimentary for up to 40,000 compute units (compartments); for 100 million compute units, the cost is \$49 per month; and for 350 million compute units, the fee is \$249 per month (Moralis billing plan May 2024). This abstract comparison highlights the difference in the total cost for documenting a car with 30,000 compartments between the two approaches, with Westerkamp's method resulting in a significantly higher cost. Gas costs are calculated based on the Ethereum price of 4,005.75\$ per ETH on March 14, 2024.

\subsection{Performance of Supply Chain Event Documentation}
The DIDChain framework documents the production, shipping, receiving and manufacturing events of the supply chain. Its efficiency in documenting supply chain events is evaluated based on the time required to record events on the underlying cheqd blockchain~\cite{cosmos_sdk}. The execution time of our framework indicates its ability to support real-time responses within supply chains. 

\begin{figure}[!ht]
    \centering
\includegraphics[width=.75\linewidth]{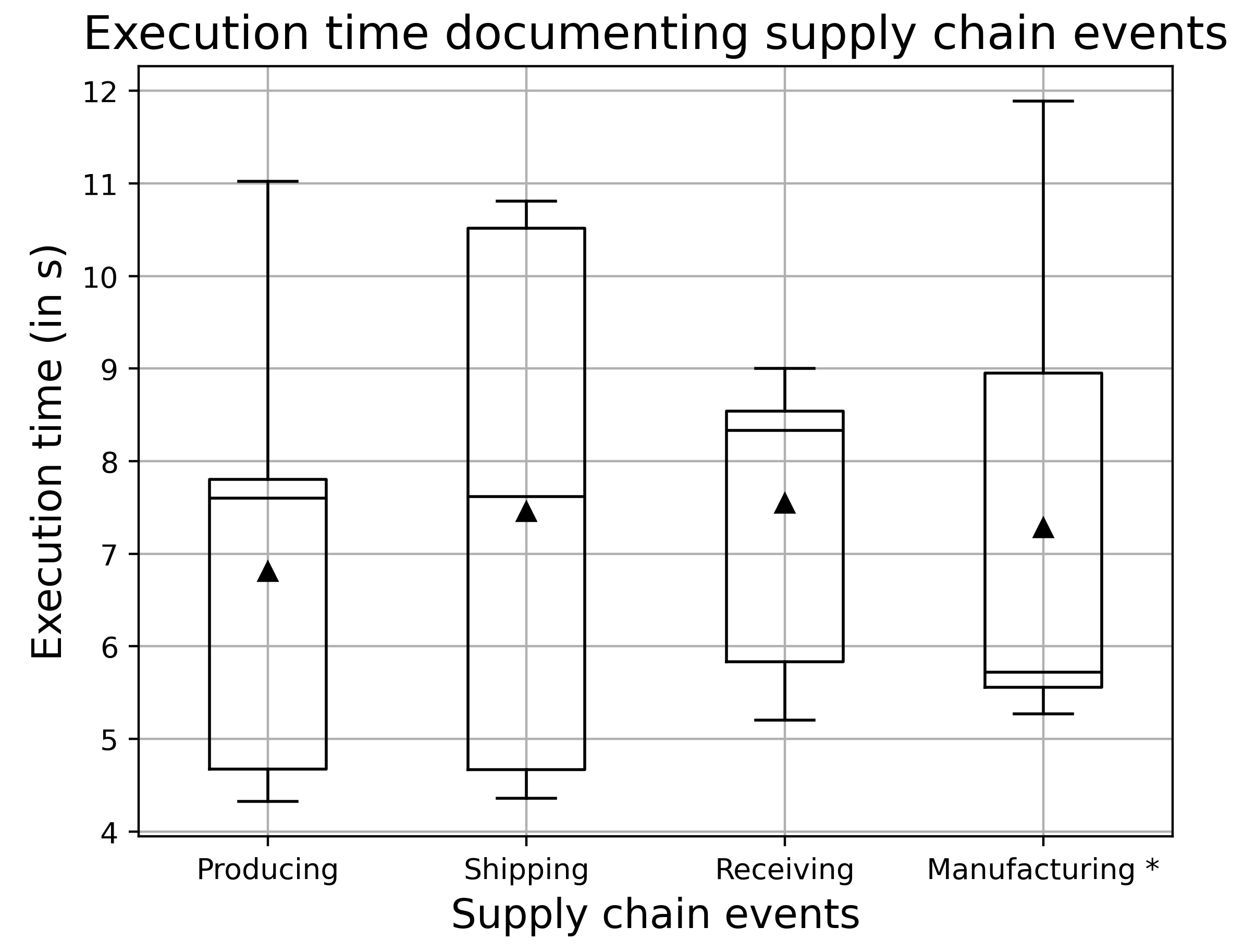}
    \caption{Comparing the average and spread of supply chain event's documentation time, tested with 30 assets per event. \** For the manufacturing event, two compartments were listed in the respective DID Document.}
    \label{fig:documentation-times}
\end{figure}

As depicted in Figure~\ref{fig:documentation-times}, the execution times for production, shipping, reception, and manufacturing events demonstrate a median documentation time that ranges from 6 to 9 seconds. The production event has a median time of 7 seconds with a standard deviation of 1.2 seconds, indicating consistent performance. Data optimization reduces documentation time by about 15\%, improving the efficiency of the system.

The interquartile range of the producing event exhibits a close concentration of data points, suggesting consistent performance with a limited deviation from the median~\cite{wan2014estimating}. The shipping event shows a broader spread, indicating more variability in execution time for this event. The manufacturing event displays occasional peaks in execution time that can affect system responsiveness. However, it maintains a median time comparable to that of the shipping event. DIDChain efficiently manages the peaks of manufacturing events, as delays in documenting manufacturing events could lead to bottlenecks in the supply chain process. The receiving and producing event shows a condensed interquartile range, suggesting that the framework can handle such events with a higher degree of predictability.

The overall distribution of an event's execution times provides insights into the system's performance and potential areas for optimization. For instance, reducing the execution time variability for shipping events could enhance the system's robustness. Furthermore, addressing manufacturing documentation time could significantly improve operational efficiency. The median execution times demonstrate the ability of the DIDChain framework to document supply chain events efficiently.

The DIDChain framework must adapt various supply chain activities with varying computational demands while maintaining a rapid and reliable documentation process. Exploring the potential of executing smart contracts and managing data more efficiently could be a promising area for research to achieve a more uniform distribution of execution times and enhance the predictability of system performance. Architectural improvements satisfy real-time SCDM systems within operational frameworks.

\subsection{Analysis of Execution Time Variability in Manufacturing}
\begin{figure}[!ht]
    \centering
\includegraphics[width=.7\linewidth]{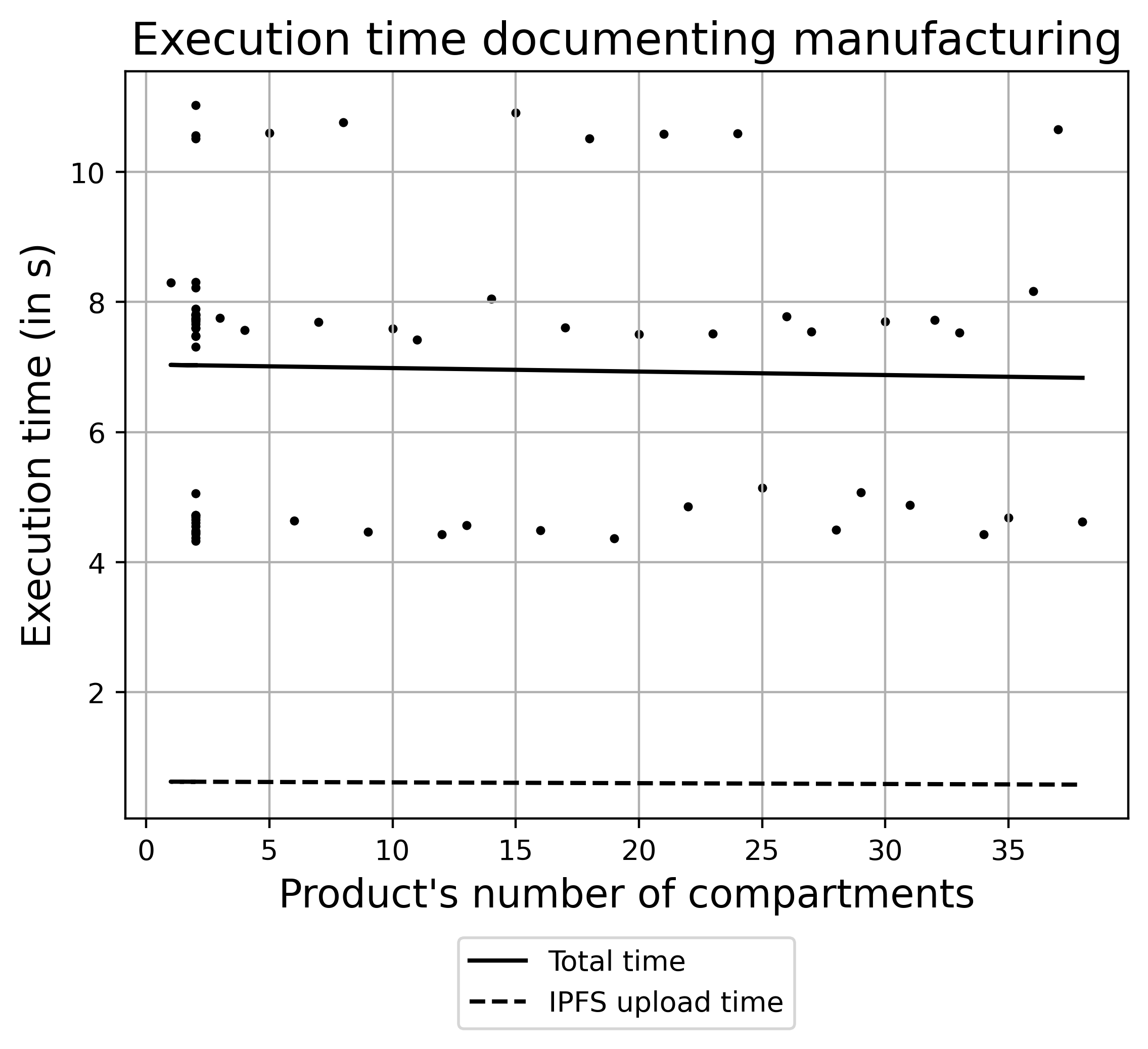}
    \caption{Comparing the performance time of documenting manufacturing of 68 products with different amounts of compartments. On average, each of these products has 11 compartments.}
    \label{fig:documentation-time-manufacturing}
\end{figure}

The number of compartments in a product affects the time required for event documentation during manufacturing processes. Figure~\ref{fig:documentation-time-manufacturing} shows the nearly constant documentation time of IPFS upload times and blockchain write transactions.

Empirical observation indicates that documenting a manufacturing event for a product ranging from one to 39 compartments takes around 7 seconds, as presented in Figure~\ref{fig:documentation-times}. However, the scope of this study is limited to products with up to 39 compartments due to observed transaction failures when attempting to assign a DID to products with 40 or more compartments. This limitation comes from the Credential Service API. The product's limitation to 39 compartments indicates a shortfall in the adaptability of the API for practical applications.
For documenting products with more than 39 compartments, an alternative strategy might involve using the Cosmos CLI cheqd node~\cite{cheqd_create_did}. 

\subsection{Constraints on DID Document Capacity}
The capacity of a product's initial DID Document is limited by the cheqd network block size threshold of $200$ KB. Each compartment adds $200$ KB to an initial DID Document baseline of $1.05$ KB so that a product can include up to 795 compartments. However, a DID Document size limit may be lower, accounting for the cheqd network's simultaneous processing of other transactions. 

\subsection{Tracing}
\begin{figure}[!ht]
    \centering
    \includegraphics[width=.8\linewidth]{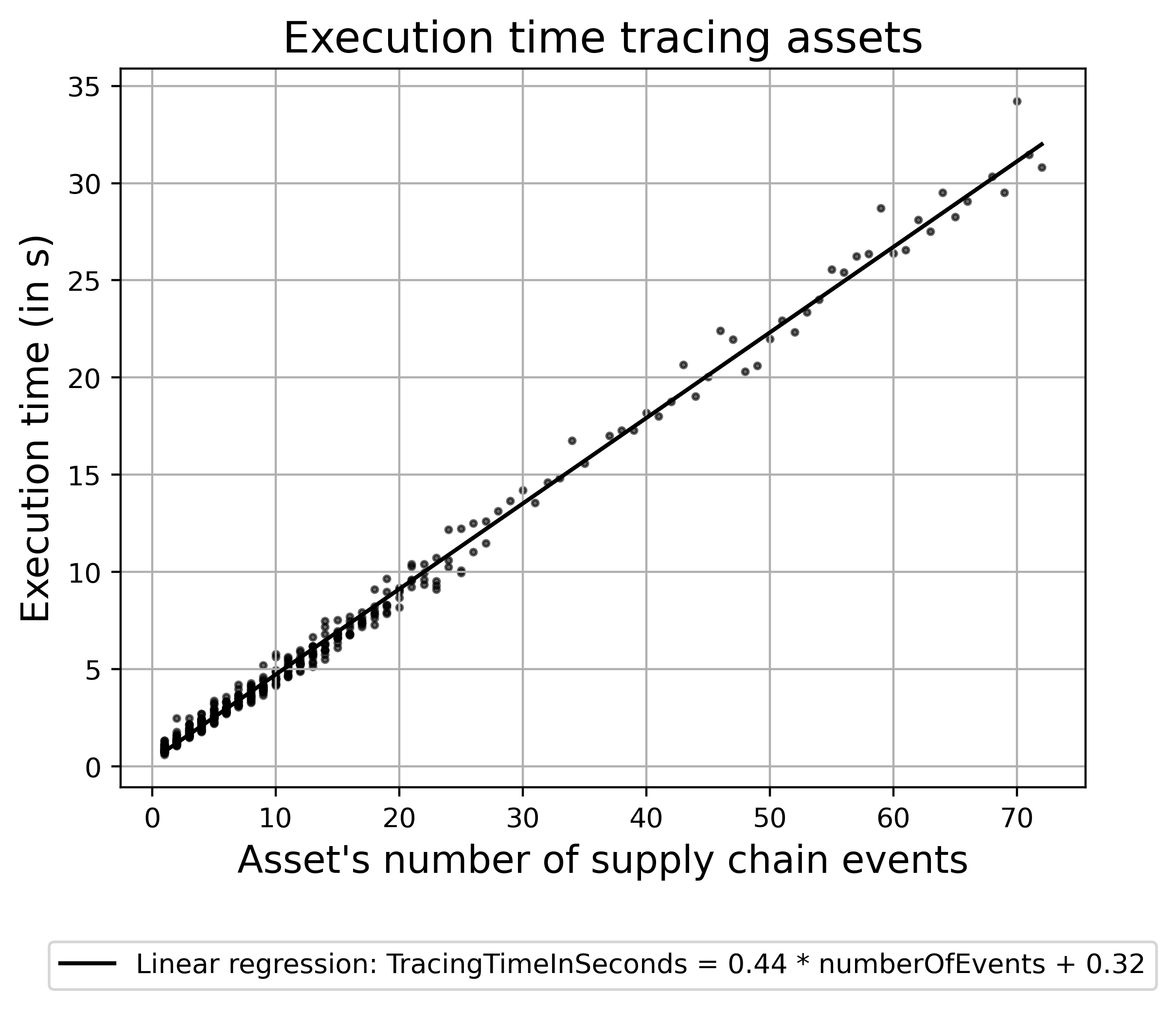}
    \caption{Comparing the performance time of tracing 383 assets with supply chains of varying lengths. On average, each asset has undergone 13 supply chain events.}
    \label{fig:execution-time-history}
\end{figure}

Tracing the history of an asset in the supply chain depends on the cumulative number of supply chain events associated with that asset. In the case of products, the history involves events directly related to the product and events related to each of its compartments. The relationship between the time required to trace an item and its number of supply chain events is illustrated as a regression graph in Figure~\ref{fig:execution-time-history}. 

The linear regression shown in Figure~\ref{fig:execution-time-history} is expressed in the following formula:
\begin{equation}
\text{TracingTimeInSeconds}(x) = a \times x + b
\end{equation}
where:
\begin{itemize}
\item \( x \) is the total number of supply chain events associated with the item,
\item \( a = 0.44 \) is the coefficient representing the time increment per supply chain event in seconds,
\item \( b = 0.32 \) is the constant representing the base time for tracing in seconds.
\end{itemize}
Function~\ref{fnc:TracingTime} represents the impact of the complexity of the supply chain and the efficiency of tracing operations. The coefficients $a = 0.44$ and $b = 0.32$ in the linear regression model obtained using the least squares estimation method, exhibit statistical significance with alpha $1\%$.
\begin{equation}
\text{TracingTimeInSeconds}(x) = 0.44 \times x + 0.32
\label{fnc:TracingTime}
\end{equation}

As supply chains become larger and incorporate more events, the time to trace the history of an item increases linearly. Therefore, the efficiency of the tracing mechanisms must be balanced with the granularity of the supply chain events to ensure the application of DIDChain in real-world scenarios. The benchmark results of our implementation of the DIDChain framework can be found here~\cite{sid030sid2024results}.

\subsection{Trust Evaluation in Decentralized Systems}
Ensuring transaction integrity and data reliability within decentralized systems such as blockchain is essential for SCDM. In the context of SCDM, our DIDChain framework fortifies trust by integrating cryptographic measures and consensus algorithms. These elements authenticate the digital assets involved and validate the authenticity of transactions, leveraging DIDs and blockchain technology.

Exploring alternatives such as producer, supplier and manufacturer authorization with OID4VC~\cite{oid4vc} and DIDComm~\cite{DIDCommV21} offers the potential to improve trust in SCDM. OID4VC introduces a method to refine issuer authentication by integrating object identifiers with the verifiable credential model, facilitating advanced security and interoperability. DIDComm promotes secure decentralized communication within the supply chain network.

\subsection{Discussion}
The evaluation of the DIDChain framework showcases its potential for SCDM by improving traceability, transparency, and trust. Despite its promising capabilities, the framework encounters challenges related to the scalability of managing complex supply chains and the economic feasibility of its deployment. The cost analysis of the DIDChain framework raises concerns about its economic viability, suggesting the need for strategic and technical modifications to ensure its broader applicability in the supply chain industry. Further investigation must assess the scalability of linked DID Documents in complex manufacturing scenarios despite the importance of DIDChain in enhancing traceability and transparency.

In response to these challenges, one potential strategy to enhance the scalability and cost effectiveness of the framework involves the integration of IPFS to store compartment lists. This approach could streamline data management by leveraging the efficiency of decentralized storage solutions. Moreover, implementing a Merkle tree~\cite{szydlo2004merkle} that consolidates the DIDs of all product compartments could be an innovative solution to facilitate the rapid and reliable verifiability of the product trace. The root of this Merkle tree, stored on the blockchain, would enable efficient verification processes while maintaining the integrity and transparency of supply chain data.

Exploring advanced DID methods, such as ION~\cite{DIF_ION}, could offer additional avenues for more economically viable strategies. These methods might provide scalable and cost-effective alternatives for DID management, potentially addressing the identified limitations of the DIDChain framework. 

Therefore, future optimizations of the DIDChain framework should focus on solving economic challenges and unlocking the full potential of blockchain technology to improve supply chain efficiency, reliability, and transparency. Our solution showed that it is not feasible in a real-world scenario and needs fundamental technical improvements.

\section{Conclusion and Future Work}
DIDChain aims to improve SCDM by integrating blockchain, IPFS, and DIDs, with potential in 6G~\cite{li2022bctrustframe}, IoT~\cite{ali2017iot}, and identity management~\cite{10246272}. DIDChain balances transparency and privacy by limiting access to sensitive data, ensuring that only authorized participants can view confidential information while maintaining traceability. Despite its potential, DIDChain faces scalability and economic feasibility challenges, necessitating further data linkage and management innovation to ensure greater applicability and efficiency.

Future research will focus on refining DIDChain by exploring the potential of smart contracts for enhanced dynamic capabilities, alongside investigating robust revocation mechanisms and verifiable credentials to improve security and trustworthiness within decentralized SCDM systems. The enhancement of the authorization of supply chain entities using the OID4VC framework will also be a key development area, aiming to solidify trust and operational integrity across the supply chain ecosystem.

Moreover, efforts to integrate DIDChain within the European Blockchain Service Infrastructure (EBSI)~\cite{EBSI2024} highlight the ambition to create a cross-border trust framework, underscoring the potential of the framework to redefine the integrity and ecosystem of digital assets. 

\section*{Acknowledgment}
This work was supported by the European Union’s Digital Europe Program (DIGITAL) research and innovation program under grant agreement number 101102743 (TRACE4EU). The authors thank the Trace4EU project partners for their invaluable contributions and the cheqd team for their support and expertise.
\bibliographystyle{IEEEtran}
\bibliography{references}

\end{document}